\documentclass[]{aa}

\usepackage{psfig}
\usepackage{graphicx}

\newcommand{\gtrsim}{\raisebox{-5pt}{$\;\stackrel{\textstyle >}{\sim}\;$}}

\begin{document}

\title{
The warm-hot intergalactic medium at $z\sim2.2$: Metal enrichment and
ionization source
\thanks{Based on observations made at the European Southern Observatory 
(ESO), under prog. ID No. 166.A-0106(A), with the UVES spectrograph 
at the VLT, Paranal, Chile.}
}

\titlerunning{}

\author{J. Bergeron\inst{1}
       \and
	B. Aracil\inst{1}
       \and
	P. Petitjean\inst{1,2}
       \and
	C. Pichon\inst{3}
}

\offprints{J. Bergeron}

\institute{
   Institut d'Astrophysique de Paris - CNRS, 98bis Boulevard Arago,
   F-75014 Paris, France
	\and
   LERMA,  Observatoire de Paris, 61 Avenue de l'Observatoire, 
   F-75014 Paris, France
	\and
   Observatoire Astronomique de Strasbourg, 11 rue de l'Universit\'e,
    F-67000 Strasbourg, France 
}

\date{Received date / Accepted date}

\abstract{Results 
are presented for our search for warm-hot gas towards the quasar 
Q 0329$-$385. We identify ten O\,{\sc vi} systems of
which two are within 5000 km s$^{-1}$ of $z_{\rm em}$ and a third one 
should be of intrinsic origin. The seven remaining systems have 
H\,{\sc i} column densities 
$10^{13.7}$$\le$$N$(H\,{\sc i})$\le$$10^{15.6}$ cm$^{-2}$. 
At least $\sim$1/3 of 
the individual O\,{\sc vi} sub-systems have temperatures 
$T$$<$1$\times$10$^5$ K and cannot originate in collisionally ionized gas.
Photoionization by a hard UV background field reproduces well the ionic 
ratios for metallicities in the range 10$^{-2.5}$-10$^{-0.5}$ solar, 
with possibly  sub-solar N/C relative abundance. 
For [O/C]=0, the sizes inferred for the O\,{\sc vi} clouds are in some cases
larger than the maximum extent implied by the Hubble flow. This constraint 
is fulfilled assuming a moderate overabundance of oxygen relative
to carbon. For a soft UV ionizing spectrum, an overabundance of O/C is
required, [O/C]$\approx$0.0-1.3. 
For a hard(soft) UV spectrum and [O/C]=0(1), 
the O\,{\sc vi} regions have overdensities 
$\rho/\overline{\rho}$$\approx$10-40.
}

\maketitle

\section{Introduction}\label{intro}


Numerical simulations suggest the existence of a warm-hot phase in the 
intergalactic medium, 
$10^5$$<$$T$$<$$10^7$ K, which comprises a fraction of the baryons increasing 
with time. This phase should be mostly driven by shocks, 
at least at low redshift $z$ (Cen \& Ostriker 
1999; Dav\'e et al. 2001). Possible signatures of the warm-hot intergalactic 
medium (WHIM) are absorptions by high ionization species
such as O\,{\sc v}, O\,{\sc vi} and O\,{\sc vii}. These absorptions are 
difficult to
detect as they either fall in the Ly$\alpha$ forest (below the atmospheric
cut-off for $z < 1.92$) or in the soft X-ray range. Successful observations of
the WHIM at low $z$ were made with the FUSE, HST and Chandra satellites 
(e.g. Tripp et al. 2001; Savage et al. 2002; Nicastro et al. 2002). 

At $z$$\sim$2-2.5, an analysis of the O\,{\sc iv}/O\,{\sc v} 
ratio from HST stacked spectra favors a hard UV background spectrum
(thus a small 
break at 4 Ryd) and the inferred metallicity is [O/H]$\simeq$$-2.2$ to $-1.3$
together with an enhanced oxygen abundance relative to carbon (Telfer 
et al. 2002). Detection of individual O\,{\sc vi} absorbers  
has been recently reported: for systems at $z$$\sim$2.5  with 
N({\rm H\,{\sc i})$\sim$$10^{14.0}$ to $10^{15.0}$ cm$^{-2}$, the inferred 
metallicity  is  [O/H]$\sim$$-3$ to $-2$ 
(Carswell et al. 2002) and for 
N({\rm H\,{\sc i})$\geq$$10^{15.5}$ cm$^{-2}$ the metallicity is higher,  
[O/H]$\geq$$-1.5$   (Simcoe et al. 2002). 
The main heating process of the high $z$ WHIM is still unclear: the more 
tenuous regions of the Ly$\alpha$ forest could be ionized by a hard UV 
background spectrum, whereas the high column density 
population could be shock heated.

  
A systematic, 
large survey of quasar absorption lines at high S/N and
high spectral resolution is being completed at ESO for a sample of about  
20 quasars of which half are at $z$$\le$2.6. 
In this paper, we present the results of our search for O\,{\sc vi} absorbers 
towards one quasar of the ESO large programme, Q~0329$-$385,  with several 
unambiguous cases of narrow, strong  and weak O\,{\sc vi} absorptions. 
The observations, the selection procedure for O\,{\sc vi} systems and our 
O\,{\sc vi} sample are presented in Sect. \ref{obser}. 
The constraints derived from the line widths are given Sect. \ref{temp}. Our
modelling of the O\,{\sc vi} absorbers  is presented in  Sect. \ref{OVI}.
The summary and conclusions are given in Sect. \ref{conc}.

\section{Observations and the O\,{\sc vi} sample}\label{obser}

The quasar Q~0329$-$385 ($z_{\rm em}$=2.423) was observed at the VLT with the 
UVES spectrograph. The full wavelength coverage 3050-10400 \AA\ was obtained
in two settings, using dichroics, with an exposure time of 6 hr per setting. 
The S/N ratio is about 30 and 100 at 3300 and 5000 \AA\ respectively.
The resolution is $b$=6.6 km s$^{-1}$. A modified version of the ESO-UVES
pipeline was used, better adapted to quasar spectra.  
A full description of the data reduction method will be presented in a 
forthcoming paper (Aracil et al. in preparation). 
The  absorption lines were fitted by multiple Voigt profiles using the VPFIT
software package  
(see http://www.ast.cam.ac.uk/{\tiny {$\sim$}}rfc/vpfit.html).

\begin{table}   
\caption[]{Column densities of the O\,{\sc vi} systems in Q 0329$-$385}
\begin{center}
\begin{tabular}{cl@{\hspace{0.5mm}}ccc@{\hspace{2.0mm}}cc}
\hline
\noalign{\smallskip}
  $z$ & n$^a$  & H\,{\sc i}&  O\,{\sc vi} & 
 N\,{\sc v}  & C\,{\sc iv} &  Si\,{\sc iv}   \\
\noalign{\smallskip}
\hline
\noalign{\smallskip}
 2.0615  &   1  &  \  14.26$^d$ &   13.13  &  $<$12.30   &   $<$12.30 & \\
 2.0764  &   1  &    13.71    &   13.18  &  \ \ \  12.89  &  \ \ \  13.23   &  
11.50$^g$  \\
 2.1470  &   3  &    14.70    &   14.03  &  $<$12.60   &  $<$12.30  &\\
 2.2488  &   5  &  \  14.00$^b$ &  14.29 &  $<$12.60  & \ \ \ \  12.83$^b$  &\\
 2.2515  &   4$^f$  &    15.59    &   14.80  & \ \ \  13.20 & 
\ \ \ \ 14.43$^e$  & 
  12.71 \\
 2.3139  &   3  &    14.23    &   13.46  &  $<$12.50 & \ \ \ \ 12.45$^c$  &\\  
 2.3521  &   1  &    13.08    &   14.04  &  \ \ \ 13.78  & \ \ \ 13.76  &\\
 2.3638  &   2  &    14.84    &   13.77  &   12.00$^g$  & \ \ \ 12.53  &\\
 2.3729  &   3  &  \  15.24$^e$ & 14.25  &  $<$12.50  & \ \ \ \ 12.69$^c$  & \\
 2.4062  &   2  &    14.03    &   13.27  &  $<$12.30   &  $<$11.70  &\\
\noalign{\smallskip}
\hline
\noalign{\smallskip}
\multicolumn{7}{l}{$^a$ Number of individual O\,{\sc vi} components.
Ions spanning}\\
\multicolumn{7}{l}{the same velocity range but with a different number of}\\
\multicolumn{7}{l}{components are marked: $^b$ $-$2, $^c$ $-$1, 
$^d$ $+$1, $^e$ $+$2.}\\
\multicolumn{7}{l}{$^f$ Main complex.}\\
\multicolumn{7}{l}{$^g$ 2$\sigma$ detection.}
\end{tabular}
\end{center}
\label{syst}
\end{table}

\begin{table}      
\caption[]{Column densities of individual O\,{\sc vi} sub-systems}
\begin{center}
\begin{tabular}{ccccccc}
\hline
\noalign{\smallskip}
$z$(O\,{\sc vi}) & H\,{\sc i} & O\,{\sc vi} & N\,{\sc v} 
& C\,{\sc iv} & Si\,{\sc iv}\\
$\left|\Delta z^a \right|$& $b$ & $b$ & $b$ & $b$ & $b$ \\
\noalign{\smallskip}
\hline
\noalign{\smallskip}
  2.24835   &  12.87 & 13.63 &   $<$12.00  &    12.21  \\
 4.1$\times10^{-5}$   &   21.8& 8.8& & 8.9  & \\
 2.25147  &  14.97 &   14.37 & \ \ \  13.00  &  14.07 &   12.43 \\
 6.0$\times10^{-5}$  & 18.0:  & 10.7  & \ \ \ 9.5  & 11.9  & 16.5  \\
 2.31395  &   13.90 &  13.20 & $<$12.30   &    11.98 \\
 0.8$\times10^{-5}$  & 29.2  & 13.4  &   & 8.6  & \\
 2.35214$^b$ &  13.08 &  14.04 & \ \ \ 13.78 & 13.76 \\
  5.7$\times10^{-5}$  & 21.8  & 12.7  & \ \ \ 9.6 &  9.1 & \\
 2.36385  & 14.55  & 13.50  & 12.00$^c$  & 12.33  & \\   
 6.3$\times10^{-5}$  & 27.5  & 9.5  &   & 10.9  & \\ 
\noalign{\smallskip}
\hline
\noalign{\smallskip}
\multicolumn{6}{l}{$^a$ Difference between
the redshifts of O\,{\sc vi} and C\,{\sc iv}.}\\
\multicolumn{6}{l}{$^b$ Intrinsic system.} \\
\multicolumn{6}{l}{$^c$ 2$\sigma$ detection.}
\end{tabular}
\end{center}
\label{indOVI}
\end{table}

\begin{figure}[!hb] \centering 
\includegraphics[width=6.4cm]{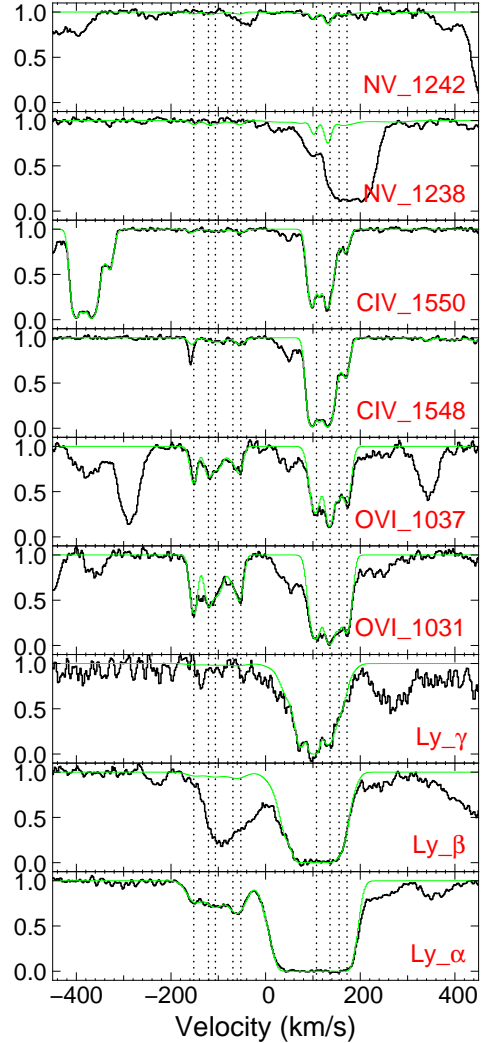}
\caption{The complex systems at $z$=2.2488 and 2.2515. The thin line shows 
the best fit model only for the lines associated with the two O\,{\sc vi} 
main clusters; there are additional components in the velocity
range between these two clusters. The line blended with 
C\,{\sc iv}1548 at $z$=2.24835 is  Mg\,{\sc i} at $z$=0.76273.}
\label{z225}
\end{figure}

\begin{figure} \centering 
\includegraphics[width=6.4cm]{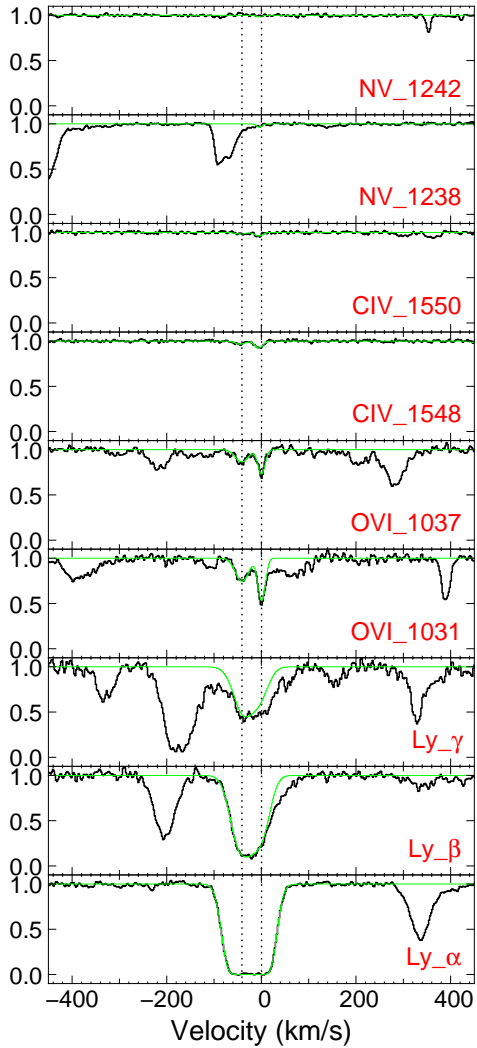}
\caption{The system at $z$=2.3638 and the best fit model. }
\label{z236}
\end{figure}


To identify the O\,{\sc vi} systems, we used the following 
criteria: either the two lines of the doublet are clearly seen
or, when  one line of the doublet is partly blended, 
the presence of this line is clearly indicated by a distinct structure 
within the blend. 
In all cases, there are H\,{\sc i} Lyman lines associated with the 
O\,{\sc vi} doublets (see Table \ref{syst}).


Our sample comprises ten O\,{\sc vi} systems of which two within 
$\Delta v$=5000 km s$^{-1}$ of the quasar redshift. The redshift,
number of components used in the fit and  total column 
densities are given in Table \ref{syst}.
The range  of $N$(H\,{\sc i}) and $N$(O\,{\sc vi}) covered by these
systems overlap with those of the samples of 
Carswell et al. (2002) and Simcoe et al. (2002).
The  3$\sigma$ detection limits for individual sub-systems are typically  
(12, 2.0 and 1.0) $\times 10^{12}$ cm$^{-2}$ for O\,{\sc vi}, N\,{\sc v} 
and C\,{\sc iv} respectively.
When  C\,{\sc iv} is detected, the upper limits for N\,{\sc v} are estimated 
using the same number of components and same b values  
as for the C\,{\sc iv} absorptions.
When neither N\,{\sc v} nor C\,{\sc iv} is detected, we assume 
$b$=10 km s$^{-1}$ and a number of components equal to those of O\,{\sc vi}.

The individual sub-systems for which a detailed modelling is presented in 
Sect. \ref{OVI} have clean O\,{\sc vi} doublets with little blending 
which ensures a correct estimate of the $b$ parameter.  
The velocity shift between the O\,{\sc vi} and  C\,{\sc iv} doublets is 
always small, $\le$5.7 km s$^{-1}$, and usually larger than that between 
the O\,{\sc vi} and Lyman lines. This was already  noted by Rollinde et 
al. (2001). The column densities and $b$ values are presented in Table 
\ref{indOVI}. The stacked velocity plots  are given only for the 
three intervening O\,{\sc vi} systems with low $b$ parameters.

\section{Temperature}\label{temp}

The histogram of the $b$(O\,{\sc vi}) values is shown in Fig. \ref{histotemp}  
for the 24 individual components of the ten systems presented in 
Table \ref{syst}, of which five are at $\Delta v$$<$5000 km s$^{-1}$ 
(shaded area).  Six (11) systems   have line widths 
$b<$10(14) km~s$^{-1}$, 
thus a temperature $T$$\leq$1.0(2.0)$\times10^5$ K. 
In the assumption of pure collisional ionization, this would imply  
6.3$\times10^{-8}$$\leq$ O\,{\sc vi}/O$\leq$2.1$\times10^{-2}$ 
(Sutherland \& Dopita 1993) and very high oxygen abundances. 
At $T$=(1.0, 1.5 and 2.0)$\times10^5$ K, the values of [O/H] are 
log(O\,{\sc vi}/H\,{\sc i})+(5.6,1.5,$-$0.7) respectively. 
For the systems and sub-systems given in Tables \ref{syst} and \ref{indOVI}
log(O\,{\sc vi}/H\,{\sc i}) is $\gtrsim$$-$1.0. Consequently, the six 
individual absorbers with $T$$\le 1.0$$\times10^5$ K would have a metallicity
[O/H]$\ge$4.6, i.e. oxygen would be the more abundant element 
log(O/H)$\ge$1.5.
This clearly rules out collisional ionization for O\,{\sc vi} absorbers with 
$b$$\le$10 km~s$^{-1}$.


In addition the
small velocity widths  constrain the size of the absorbers. At 
$z$$\sim$2.2, the line broadening due to the Hubble flow implies a maximun 
pathlength $l_{\rm H}$=80 $b_{10}$ kpc, assuming  
H$_0$=65 km s$^{-1}$ Mpc$^{-1}$,
$\Omega_{\rm M}$=0.3 and $\Omega_{\rm \Lambda}$=0.7. 
In Sect. \ref{sub}, this value is compared to the size of the absorbers 
based on the gas density derived from photoionization models in order 
to constrain the spectral energy distribution (SED) of the UV background 
spectrum.

\begin{figure} \centering 
\includegraphics[width=4.5cm]{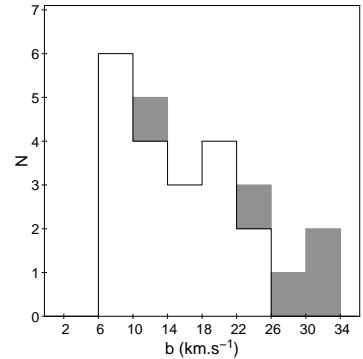}
\caption{Histogram of the b values of individual sub-systems. }
\label{histotemp}
\end{figure}

\section{Analysis of the O\,{\sc vi} systems}\label{OVI}

Different SEDs were used for the UV background radiation at $z \sim 2.2$. 
The first one is a simple analytic  
fit to the spectrum given by Haardt \& Madau (1996: hereafter HM96) with an 
energy spectral index  $\alpha$ = 1.5 ($f_\nu \propto \nu^{-\alpha}$) 
shortwards of 1 Ryd, a break of $10^{0.5}$ at 4 Ryd and adopting  
their normalization of the ionizing continuum at 1 Ryd. We also consider
the spectrum derived by Madau et al. (1999: hereafter MHR99) which is 
softer with  a stronger break ($10^{1.4}$) at 4 Ryd. 
Note that the analyses of  He\,{\sc ii}/H\,{\sc i} (Kriss et al. 2001),
 O\,{\sc iv}/O\,{\sc v} (Telfer et al. 2002) and 
O\,{\sc vi}/C\,{\sc iv}  (Carswell et al. 2002)
 suggest a hard ionizing spectrum similar to that of HM96.

\subsection{The Oxygen abundance}\label{O}

We used the CLOUDY code  (Ferland et al. 1998) to derive the ionization 
levels of the different species and their column density ratios. 
Results are given for optically thin systems, [C/H]=$-1$ and solar 
relative abundances. The ionization balance is very similar for any 
other values of [C/H]$<$$-1$ as the temperature, thus ionic ratios, 
is not strongly dependent on metallicity. 
Fig. \ref{abondances} gives  
$N$(O\,{\sc vi})/$N$(H\,{\sc i}) $\equiv$ O\,{\sc vi}/H\,{\sc i} as a function 
of $N$(O\,{\sc vi})/$N$(C\,{\sc iv}) $\equiv$ O\,{\sc vi}/C\,{\sc iv}. 
Variation of the metallicity results 
in a vertical shift of the theoretical curves.
One system ($z$= 2.3521  
and O\,{\sc vi}/H\,{\sc i}=9.1) is clearly different from the 
rest of the absorbers: its metallicity should be about solar which 
strongly suggests an intrinsic origin. The presence of strong
N\,{\sc v} associated absorption reinforces this conclusion.
For the other systems, their positions relative to the theoretical 
curves imply metallicities in the range   
[C/H]$\approx$$-0.5$ to $-2.5$ for the harder ionizing spectrum.   

\begin{figure}[!ht] \centering 
\includegraphics[width=6.0cm]{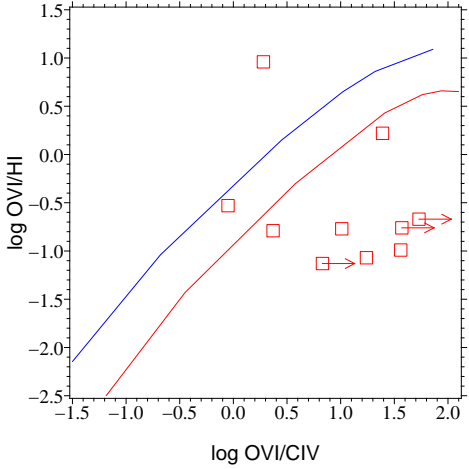}
\caption{The O\,{\sc vi}/H\,{\sc i} ratio versus  
O\,{\sc vi}/C\,{\sc iv}. The red and blue 
lines show the ratios obtained with the HM96 and MHR99 spectra respectively,
 [C/H]=$-$1.0 and solar relative abundances.
 The squares are the observed values for the ten systems given in 
Table \ref{syst} and upper limits are indicated by arrows.}
\label{abondances}
\end{figure}

\begin{figure} \centering 
\includegraphics[width=6.0cm]{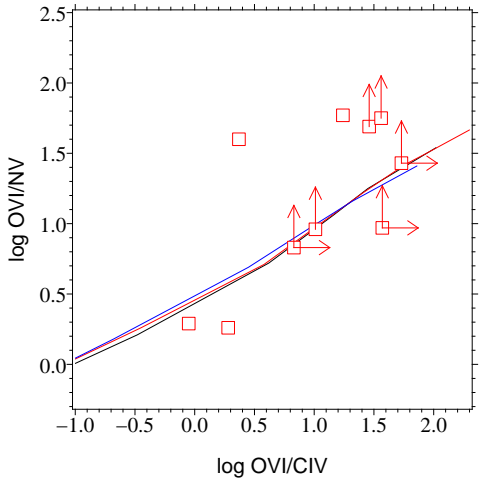}
\caption{The O\,{\sc vi}/N\,{\sc v} ionic ratio versus that  of 
O\,{\sc vi}/C\,{\sc iv}. The red and blue lines 
refer to the same models as in Fig. \ref{abondances} and, for the
black one, the spectral slope $\alpha$ = 1.5 and the break at 4 Ryd equals
 $10^{1.0}$. The other symbols are as in Fig. \ref{abondances}.}
\label{rapports}
\end{figure}

The column density ratios of high ions is shown in Fig.~\ref{rapports}. 
There is a single, well defined locus for the theoretical models with 
three different SEDs (HM96, MHR99 and our simple analytic fit to the HM96 
spectrum with a break of $10^{1.0}$) and solar relative abundances. 
Two systems, with detected O\,{\sc vi}, N\,{\sc v} and C\,{\sc iv} doublets,
lie below this locus. The furthest
away is the above mentioned, intrinsic system: the inferred sub-solar [O/N]
abundance ratio is indeed consistent with the elemental abundances 
(enhanced nitrogen abundance relative to solar ratios) 
typical of most intrinsic absorbers (Hamann 1997). 
For four out of the other five  systems with detected 
O\,{\sc vi} and C\,{\sc iv} doublets, the observed ionic ratios imply either 
a super-solar [O/C] abundance ratio, as also found by Telfer et al. (2002)
for WHIM clouds, and/or a sub-solar [N/C] abundance ratio.

\begin{figure} \centering 
\includegraphics[width=6.0cm]{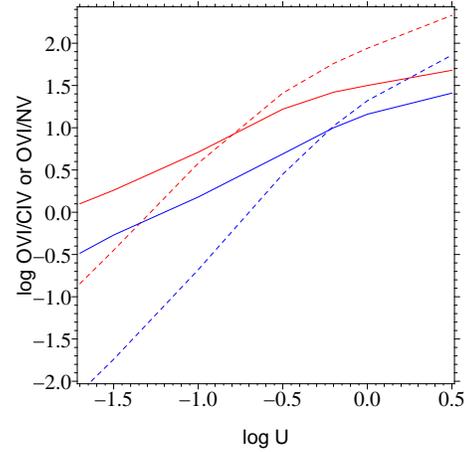}
\caption{The O\,{\sc vi}/C\,{\sc iv} (full lines) or O\,{\sc vi}/N\,{\sc v} 
(dotted lines) ionic ratio versus the ionization parameter U.  
Red and blue lines refer to the HM96 and MHR99 spectrum respectively 
together with solar relative abundances.}
\label{U}
\end{figure}

\subsection{Pathlength and abundances of some individual
sub-systems}\label{sub}

For each system the ionization parameter, U, is determined from  
O\,{\sc vi}/C\,{\sc iv} adopting a HM96 spectrum and solar relative 
abundances (Model A: see Fig. \ref{U} and Table \ref{Ul}); the pathlength  
is derived from the modelled physical state of the gas.
For the three sub-systems with O\,{\sc vi}/H\,{\sc i}$>$0.1, one with 
$N$(H\,{\sc i})$>$1$\times10^{15}$ cm$^{-2}$,
the inferred sizes are compatible with limits given by the Doppler
parameters (see Sect. \ref{temp})
for a HM96 ionizing spectrum and solar relative abundances;  
the temperature is $T$=(3.0, 3.0 and 4.8)$\times10^4$~K 
for the $z$=2.24835, 2.25147 and 2.31395 absorbers respectively. 
However, for the $z$=2.36385 sub-system with  O\,{\sc vi}/H\,{\sc i}$<$0.1, 
the inferred size is a factor $\sim$2 larger than the maximun size due to 
Hubble expansion; to have $l/l_{\rm H}$$\leq$1 implies [O/C]$\gtrsim$0.3.

A second model is then investigated assuming a HM96 spectrum and 
[O/C]=1.0  (Model B). The derived ionization parameters are $\sim$3 times 
lower and $l$$\ll$$l_{\rm H}$ for all the individual sub-systems. 
Finally, since softer spectra imply higher U values, thus lower densities 
and larger pathlengths, we consider a MHR99 spectrum only together with  
[O/C]=1.0 (Model C). Results for Models A and C are very similar as the 
decrease in U due to the enhanced [O/C] relative abundance (see 
Sect. \ref{O}) is compensated by the increase in U required by the larger 
break at 4 Ryd  in the softer spectrum.
The range in metallicity of the WHIM absorbers presented in Table~\ref{Ul}
is [C/H]$\sim$$-0.7$ to $-2.9$.


The constraints on the SED of the background ionizing spectrum and the [O/C] 
relative abundance are thus linked. Even for an ionizing spectrum as hard as 
the HM96 one, we find that there must be some O\,{\sc vi} clouds with a 
super-solar [O/C] relative abundance. Adopting the MHR99 soft spectrum 
implies  [O/C]$\approx$0.0 to 1.3.

Note that in the $z$=2.25147 absorber, a fraction of  C\,{\sc iv} could 
reside  in the Si\,{\sc iv} phase. 
Using O\,{\sc vi}/N\,{\sc v} instead to derive U, with [O/N]=0
and adopting the HM96 spectrum, gives $l$$\simeq$1 Mpc, inconsistent 
with the Hubble expansion constraint. The latter is fulfilled for 
[O/N]$\geq$0.7.

\begin{table}      
\caption[]{Abundances and pathlength of the O\,{\sc vi} absorbers}
\begin{center}
\begin{tabular}{ll@{\hspace{0.5mm}}r@{\hspace{3.0mm}}c@{\hspace{2.0mm}}r@{\hspace{1.5mm}}r@{\hspace{2.0mm}}c}
\hline
\noalign{\smallskip}
 $z$(O\,{\sc vi})  &  M$^a$ &   [N/C] &  U & H/H\,{\sc i} & 
$l$(kpc)$^b$ &   [C/H]  \\
\noalign{\smallskip}
\hline
\noalign{\smallskip}
 2.24835 &     A   &   $<$--0.43 &  $ -$0.50  &  5.21  &  6.5  & $-0.67$  \\
&     B  &   $<$--0.05  &   $-$1.05  &     4.58  &  0.4  &   $-0.84$  \\
&      C  &   $<$0.03  &   $-$0.50  &     5.21  &   6.5  &  $-1.39$  \\ 
 2.25147 &   A    &  $-0.76$  &  $-$1.08  &     4.55  &  47.0  & $-1.13$  \\
&     B  &   $ -0.18$  &  $ -$1.60  &  3.98  &  3.8  &  $-0.90$  \\
&      C  &   $ -0.22$  &  $ -$1.05  &     4.58  &  54.0  &   $-1.45$  \\
 2.31395 &     A  &   $<$0.21  &  $ -$0.63  &   5.07  &   37.0  &  $-2.00 $ \\
&     B  &   $<$0.63  &  $-$1.18  &     4.45  & 2.5  &    $-2.00$  \\
&   C  &   $<$0.68  &   $-$0.60  &     5.10  &   43.0  &   $-2.65$  \\
 2.36385 &  A  &  $\sim$--0.45  & $-$0.66  &     5.03  & 141.0  &   $-2.31$  \\
&    B  &   $\sim$0.01  &  $-$1.22  &   4.41  & 9.3  &   $-2.28$  \\
&    C  &   $\sim$0.02  &  $-$0.64  &   5.05  &  155.0   &  $-2.92 $ \\
\noalign{\smallskip}
\hline
\noalign{\smallskip}
\multicolumn{7}{l}{$^a$ Model A: $\alpha$=1.5, log (He\,{\sc ii}
break)=0.5, [O/C]=0.0,}\\
\multicolumn{7}{l}{Model B: $\alpha$=1.5, log (He\,{\sc ii} break)=0.5, 
[O/C]=1.0,}\\
\multicolumn{7}{l}{Model C: Madau et al. (1999), [O/C]=1.0}\\
\multicolumn{7}{l}{$^b$ Hubble flow constraint: 
$l_{\rm H}$$\leq$80 $b_{10}$ kpc.}
\end{tabular}
\end{center}
\label{Ul}
\end{table}

\section{Summary and conclusions}\label{conc}

A sample of ten O\,{\sc vi} absorption systems at $z$=2.06-2.41 was
identified in Q 0329$-$385. Seven systems  trace the 
warm-hot intergalactic medium, outside the close neighbourhood of the quasar. 
All the detected intervening O\,{\sc vi} systems have associated H\,{\sc i} 
absorption with column densities in the range $10^{13.7}$-$10^{15.6}$ 
cm$^{-2}$. 
At least $\sim$1/3 of the individual O\,{\sc vi} subcomponents
have low velocity dispersions, $b$$<$10 km s$^{-1}$. These unambiguous cases 
of low temperature, $T$$<$1$\times 10^5$ K, imply a radiative ionizing process
for the teneous regions of the WHIM.

The observed O\,{\sc vi}/H\,{\sc i} and  O\,{\sc vi}/C\,{\sc iv} ionic ratios
are well reproduced by photoionization models with a hard (HM96) spectrum
and [O/C]=0, whereas the weakness or non detection of N\,{\sc v} implies 
[O/N]$>$0 in most cases. The range in metallicity for our small sample is
large, [C/H]$\approx$$-$0.5 to $-$2.5, which strongly suggests a non-uniform 
enrichment of the intergalactic medium. 
The size inferred for low $b$ absorbers of smaller O\,{\sc vi}/H\,{\sc i}
ratios are about twice as large as that inferred from the Hubble flow 
constraint. For these absorbers, higher gas densities, thus sizes 
compatible with the Hubble flow, are obtained for [O/C]$\sim$0.3.  
 
An enhanced oxygen abundance is nearly always required for a softer spectrum 
(i.e. a larger break at 4 ryd), with [O/C]$\approx$0.0-1.3,  
and the enrichment of the WHIM should then comes 
predominantly from massive stars. Adopting the MHR99 spectrum  and 
[O/C]=1 leads to a metallicity range [C/H]$\approx$$-$1.2 to $-$3.0. 

The $N$(H\,{\sc i}) range of the O\,{\sc vi} systems in Q 0329$-$385 is
similar to that of the O\,{\sc vi} sample of Carswell et al. (2002) and these
authors also favor photoionization of the WHIM. They rule out a soft UV flux
 for a solar [O/C] abundance ratio, as also found for most of the O\,{\sc vi}
systems in Q 0329$-$385 but, as discussed in Sect. \ref{sub}, 
a softer UV flux is acceptable if [O/C] is super-solar.
On the contrary, Simcoe et al. (2002)  favor 
shock-heated gas. However, their O\,{\sc vi} survey  does not trace   
regions of lower column densities (no system with 
$N$(H\,{\sc i})$<$1$\times10^{14.5}$ cm$^{-2}$)  and five 
out of their 12 intervening O\,{\sc vi} absorbers have very large 
$N$(H\,{\sc i}) ($10^{16.9}$ to $10^{19.5}$ cm$^{-2}$).  
The O\,{\sc vi} systems with stronger H\,{\sc i} absorption 
may trace the  outer parts of high $z$ galatic halos 
where the effects of galactic winds may dominate the heating process
(Theuns et al. 2002), one of the models also suggested by Simcoe et al. 
(2002).

For a hard UV spectrum and [O/C]=0, the gas density is in the range 
n$_{\rm H}$=(7-23)$\times10^{-5}$ cm$^{-3}$,  thus O\,{\sc vi} arises in   
regions with overdensities $\rho/\overline{\rho}$$\approx$10-40 
for $\Omega_{\rm b} h^2_{100}$=0.02 (O'Meara et al. 2001).
Adopting a MHR99 soft spectrum and [O/C]=1 leads to the same range of 
 overdensities, whereas for the HM96 spectrum and [O/C]=1 the  
overdensities are about three times larger.

We are currently building a large  O\,{\sc vi} sample from the dataset of 
the VLT-UVES quasar large programme to investigate the physical state and 
chemical evolution of the O\,{\sc vi} absorber populations, as well as 
the relative occurence of O\,{\sc vi} and C\,{\sc iv} systems.


\begin{acknowledgements}
We are grateful to R. Carswell for stimulating discussions and his assistance
with the VPFIT s/w code.
\end{acknowledgements}





\end{document}